\documentclass[epj]{svjour}
\usepackage{graphicx,amsmath,amssymb,bm,xcolor,tabularx}
\usepackage{hyperref}
\begin{document}
\begin{sloppypar}
\title{Description of nuclear photoexcitation by Lorentzian  expressions for electric dipole photon strength function}
\author{Vladimir Plujko\inst{1,2,}
\thanks{\emph{Corresponding author:} plujko@gmail.com}%
, Oleksandr Gorbachenko\inst{1} \and Kateryna Solodovnyk\inst{1}
}                     
%
%

\institute{Taras Shevchenko National University of Kyiv, Volodymyrska str.60, Kyiv 01601, Ukraine \and
Institute for Nuclear Research, Prosp. Nauki 47, Kyiv 03028, Ukraine}

\date{Received: date / Revised version: date}
%
\abstract{
The description of photoabsorption cross-sections of cold nuclei by closed-form Lorentzian models of photon strength functions for photoexcitation by electric dipole gamma-rays is considered. Systematics of the GDR parameters are given and input parameters of different analytical models are discussed The experimental data are compared with theoretical calculations for even-even nuclei using criteria of minimum of both least-square value and root-mean-square deviation factor. Simple extensions of the models with energy-dependent widths to high gamma-ray energies $\gtrsim $ 30MeV which hold  the energy-weighted sum rule for E1 gamma-transitions in good approximation  are proposed and tested.
} 
%

\maketitle
\section{Introduction}
\label{intro}

Photon strength functions (PSF) are an important constituent of the calculations of various decay properties of the atomic nuclei, for example, the average probabilities of $\gamma $-transitions, the gamma-ray capture cross-sections, gamma-ray production spectra, isomeric state populations, competition between gamma -- ray and particle emission, as well as  different nuclear structure characteristics (deformations, contribution of velocity-dependent forces, shape-transitions, etc) (see, for example, \cite{RIPL,RIPL2,RIPL3} for references). The photon strength functions are mandatory component of all modern computer codes for nuclear reaction calculations and nuclear data evaluation, such as \cite{Talys,EMPIRE}. Some applications, such as nuclear astrophysics, also require the determination of radiative neutron data for a large number of exotic nuclei, and for energies that cannot be reached in the laboratory  \cite{Arn2007}. In this case, large-scale calculations need to be performed on the basis of sound and accurate models to ensure a reliable extrapolation far away from the experimentally known region.

The photonuclear data are being intensively studying in response to their growing needs in a variety of practical applications. For example, bremsstrahlung radiation by electron accelerators is now used in many laboratories, industries and hospitals dealing with activation analyses, radiation shielding and radiation transport analyses, calculation of absorbed dose in the human body during radiotherapy etc (see, for example, \cite{Gor2017} for references).

The electric dipole (E1) gamma-transitions are dominant in nuclear processes of photoabsorption and gamma-decay when they occur simultaneously with transitions of other multipolarities. For middle-weight and heavy atomic nuclei, the most important contribution to probability of these transitions in the range  of gamma-ray energies
$8 < \varepsilon _{\gamma } < 30$ MeV is resulted from response of  Isovector Giant Dipole Resonances (GDR). A Lorentz shape is preferable for approximation of a strength function of such response (see \cite{Plu2018} and refs therein). The common-used analytical models with this line-shape are the Standard Lorentzian (SLO), the Generalized Lorentzian (GLO), the Simplified version of  Modified Lorentzian (SMLO) \cite{RIPL,RIPL2,RIPL3,Plu2018,Plu2011,Gor2019}, the Triple Lorentzian (TLO) \cite{Jun2008,Gro2017}.

The comparisons  of  experimental data on photoabsorption cross-section with theoretical calculations within framework of  SLO and SMLO approach were done for 162 nuclei in ref.\cite{Plu2018}. It was shown that the low-energy tails of the photoabsorption cross sections within SLO model are, as a rule, higher than for the SMLO model and the experimental data; that is low-energy wings  of GDR PSF component can not be described in SLO model without re-adjusting standard values of its GDR parameters that provided description of the GDR  range of the photoabsorption cross-sections.

 The calculations of E1 PSF within microscopic both quasiparticle random-phase approximation (QRPA) and shell-model (SM) were compared with the SMLO strengths in ref.\cite{Gor2019} and had been shown that a rather good agreement was obtained. These results for E1 PSF together with updated expression for dipole magnetic (M1) strength allowed to perform relatively good description of all available experimental data on average radiative widths of the E1+M1 transitions.

In this contribution the closed-form Lorentzian models of SLO, GLO, SMLO, TLO  are tested for description of the experimental photoabsorption cross sections  in cold even-even atomic nuclei from EXFOR data-base \cite{EXFOR}  in the range of the gamma-rays till $\sim$ 30 MeV.  The  data are compared with theoretical calculations using for the merit functions  the  least-square deviation and the root-mean-square (rms) deviation factor. The extended SMLO and GLO models with saturated energy-dependent widths above the GDR energy are proposed. For these models, energy-weighted sum rule for E1 gamma-ray photoabsorption is approximately fulfilled.

\section{Photoabsorption cross-sections and the analytical expressions for E1 PSF of photoabsorption  of atomic nuclei in ground states}
\label{sec:1}

The theoretical total photoabsorption cross sections $\sigma _{{\rm abs}} (\varepsilon _{\gamma } )$ of the gamma-ray with energy $\varepsilon _{\gamma } $ are calculated as a sum of the terms corresponding to the GDR excitation ($\sigma _{\rm{GDR}} $) and quasi-deuteron photodisintegration (a photoabsorption cross section on a neutron-proton pair, $\sigma _{\rm{qd}} (\varepsilon _{\gamma } )$):

\begin{equation}
\label{EQ_1}
\sigma _{\rm{abs}} (\varepsilon _{\gamma } )=\sigma _{\rm{GDR}} (\varepsilon _{\gamma } )+\sigma _{\rm{qd}} (\varepsilon _{\gamma } ),
\end{equation}

\noindent where the approach from ref.\cite{Cha1991,Cha2000} is used for a quasi-deuteron contribution.

The total photoabsorption cross section is adopted to be equal to the photoabsorption cross-section of electric dipole gamma-rays summed over states with all possible total angular momentum $\sigma _{\rm{E1}} (\varepsilon _{\gamma } )$, because we will use experimental data for nuclear photoexcitation induced by bremsstrahlung which is mainly characterized by E1 multipolarity \cite{Cha2000}.

Component $\sigma _{\rm{ GDR}} $ of total photoabsorption  cross-section (\ref{EQ_1}) resulted from the GDR excitation is calculated by the use of different analytical models for E1 PSF of photoexcitation $\overrightarrow{f}^{\alpha } (\varepsilon _{\gamma } )$ \cite{RIPL,RIPL2,RIPL3,Bar1973}:

\begin{equation}
\label{EQ_2}
\sigma _{\rm{GDR}} (\varepsilon _{\gamma } )=\sigma _{\rm{GDR}}^{\alpha } (\varepsilon _{\gamma } )=3\left(\pi \hbar c\right)^{2} \cdot \varepsilon _{\gamma }\cdot \overrightarrow{f}^{\alpha } (\varepsilon _{\gamma } ),
\end{equation}

\noindent where index $\alpha $ denotes  PSF model.

 General analytical expression for E1 PSF of photoexcitation of cold nuclei governed by the GDR can be presented in  the following form

\begin{equation}
\label{EQ_3}
\begin{array}{l}
{\displaystyle\overrightarrow{f}^{\alpha } (\varepsilon _{\gamma } )=\frac{1}{3\cdot (\pi \hbar c)^{2} } \sum _{j=1}^{j_{m} }\sigma _{\rm{TRK}} s_{j}^{\alpha } \frac{F_{j}^{\alpha } (\varepsilon _{\gamma } )}{\varepsilon _{\gamma } }  =} \\
{\displaystyle\hspace{0.3 cm} ={\rm 8.674}\cdot 10^{-8} \, \sum _{j=1}^{j_{m} }\sigma _{\rm{TRK}} [{\rm mb}\cdot {\rm MeV}]\times} \\
{\displaystyle\hspace{0.3 cm} \times s_{j}^{\alpha } \frac{F_{j}^{\alpha } (\varepsilon _{\gamma } )[{\rm MeV}^{-1} ]}{\varepsilon _{\gamma } [{\rm MeV}]}  \ \ {\rm MeV}^{-3}.}
\end{array}
\end{equation}

Here, index $j$ numbers the normal modes of giant vibrations: $j_{m} =1$ for spherical nuclei, $j_{m} =2$ for  axially symmetric ones, and $j_{m} =3$ for nuclei with triaxial shape; factor $s_{j}^{\alpha } $ is a weight of the \textit{j}-mode; $\sigma_{\rm{TRK}}$ is the Thomas-Reiche-Kuhn (TRK) sum rule

\begin{equation}
\label{EQ_4}
\sigma _{\rm{TRK}} =60\frac{NZ}{A} =15A(1-I^{2} ) \ \ {\rm mb}\cdot {\rm MeV}
\end{equation}

\noindent  with $I=(N-Z)/A$ for the neutron-proton asymmetry factor. A weight of the \textit{j}-mode determines cross-section $\sigma _{r,j}^{\alpha } =(2/\pi )\sigma _{\rm{TRK}} \cdot s_{j}^{\alpha } /\Gamma _{j}^{\alpha } $ of $j$ -mode at resonance energy $E_{r,j}^{\alpha } $.

The line-shape functions $F_{j}^{\alpha } $ are described by Lorentzian,

\begin{equation}
\label{EQ_5}
F_{j}^{\alpha } (\varepsilon _{\gamma } )=\frac{2}{\pi } \frac{\, \varepsilon _{\gamma }^{2} \, \Gamma _{j}^{\alpha } }{(\varepsilon _{\gamma }^{2} -(E_{r,j}^{\alpha } )^{2} )^{2} +(\Gamma _{j} ^{\alpha } \varepsilon _{\gamma } )^{2} },
\end{equation}

\noindent but with different determination of the parameters of energy $E_{r,j}^{\alpha } $ and shape width $\Gamma _{j}^{\alpha } $.

An approximation of axially symmetric nuclei with the effective quadrupole deformation parameter $\beta _{2} $ is adopted for deformed nuclei in the SLO, GLO and SMLO models. The values of effective quadrupole deformation parameters are taken from ``deflib.dat'' file of RIPL 2 ($\beta _{2} $=$\beta _{2,eff} $) \cite{RIPL2}. The indexes \textit{j=}1, 2 denotes normal modes of vibrations with low  and high values of the ``resonance'' energy $E_{r,j}^{\alpha } $: $E_{r,1}^{\alpha } $$<$$E_{r,\, 2}^{\alpha } $.

For the SLO model, the $\Gamma _{j}^{\rm{SLO}} $ is energy-independent  constant which is equal to GDR width for  \textit{j}-mode and the energy $E_{r,j}^{\rm{SLO}} $ is equal to the GDR energy.  The $\Gamma _{j}^{\rm{GLO}}$ in GLO model is  quadratic in gamma-ray energy\cite{RIPL,RIPL2,RIPL3}:

\begin{equation}
\label{EQ_6}
\Gamma _{j}^{\rm{GLO}} (\varepsilon _{\gamma } )=\frac{\Gamma _{r,j}^{\rm{SLO}} }{(E_{r,j}^{\rm{SLO}} )^{2} } \cdot \varepsilon _{\gamma } ^{2}.
\end{equation}

The width in the SMLO model is a linear function of the gamma-ray energy:

\begin{equation}
\label{EQ_7}
\Gamma _{j}^{\rm{SMLO}} (\varepsilon _{\gamma } )=\frac{\Gamma _{r,j}^{\rm{SMLO}} }{E_{r,j}^{\rm{SMLO}} } \cdot \varepsilon _{\gamma }.
\end{equation}

For GDR characteristics in the SLO, GLO and SMLO, the recommended values (the energies, widths and weights)  from recent data-base \cite{Plu2018} were used and the characteristics of SLO model were taken for GLO parameters.
In absence of the experimental data values of GDR parameters by their systematic values are taken. The systematics of the energies $E_{r,j}^{\alpha }$ were obtained by the least-square fitting the recommended experimental GDR parameters in spherical nuclei and in deformed nuclei from mass-number ranges 150$<A<$190 and 220$<A<$253 where they can be considered as axially deformed ones.

These systematics  were based on simultaneous fitting both resonance energy in spherical nuclei and average resonance  energy $E_{r}^{\alpha}$  in an axially deformed nuclei \cite{Ber1975,DiT1986},

\begin{equation}
\label{EQ_8}
E_{r}^{\alpha}=\left(s_{a}^{\alpha}\cdot E_{a}^{\alpha}+s_{b}^{\alpha} \cdot E_{b}^{\alpha} \right)/s^{\alpha }, \ \ s^{\alpha}=s_{a}^{\alpha} +s_{b}^{\alpha},
\end{equation}

\noindent where $E_{a}^{\alpha}$ ($E_{b}^{\alpha}$) are the energy of the vibration along (perpendicular) to the symmetry axis and $s_{a}^{\alpha}$ ($s_{b}^{\alpha }$) is corresponding weight.

\begin{figure}[htbp]
    \includegraphics[width=0.95\columnwidth]{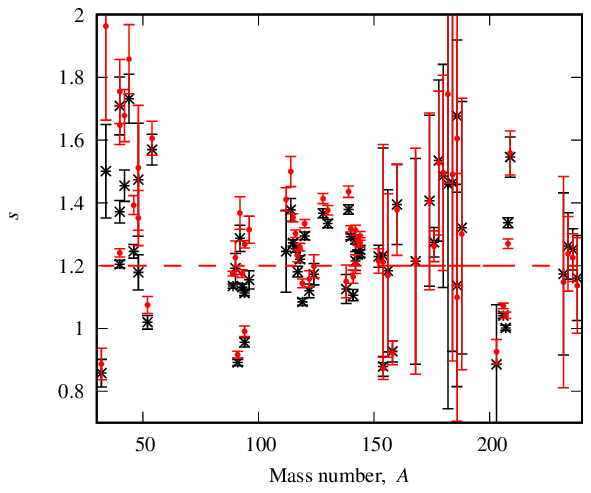}
    \includegraphics[width=0.95\columnwidth]{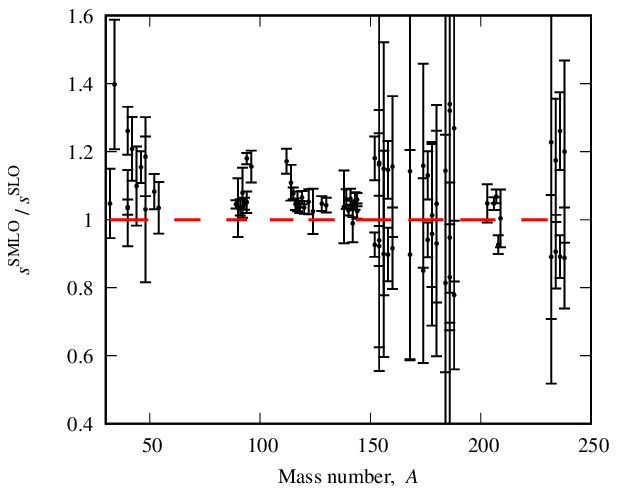}
	\caption{Experimental values of the weights $s^{\alpha } $=$s_{a}^{\alpha } $+$s_{b}^{\alpha } $ within SMLO and SLO approaches.
     Upper panel: SMLO values - red full circles with errorbars; SLO values -black stars with the errorbars;  Bottom  panel - ratios  $s^{\rm{SMLO}} $/$s^{\rm{SLO}}$.}
	\label{fig_1}
\end{figure}

In fig.\ref{fig_1} the comparison of the experimental values for a sum of the weights  for SLO and SMLO models with uncertainties from ref.\cite{Plu2018} are presented in spherical nuclei and in deformed nuclei with 150$<A<$190 and 220$<A<$253 where they can be considered as axially deformed ones. It can be seen that the systematical value
$s^{\alpha } $=1.2 can be taken for a sum of the weights  for SLO and SMLO models. For GLO model the values $s_{j}^{\rm{GLO}} =s_{j}^{\rm{SLO}}$ were adopted.

The approximation of equally probable excitation of normal modes of the giant collective vibration was taken  for systematics of the weights,

\begin{equation}
\label{EQ_9}
\left\{\begin{array}{l} {s_{1}^{\alpha } =s^{\alpha } /3;\, \, \, \, \, \, \, \, \, \, s_{2}^{\alpha } =2s^{\alpha } /3,\, \, \beta _{2} >0,} \\
{s_{1}^{\alpha } =2s^{\alpha } /3;\, \, \, \, \, \, \, s_{2}^{\alpha } =\, \, s^{\alpha } /3,\, \, \, \, \beta _{2} <0,} \end{array}\right.  \end{equation}

\noindent  because of twofold degeneration of the giant collective vibration, which is perpendicular to the axis of symmetry.

The following general expression was adopted for systematics of the GDR energy in spherical nuclei and for average energy in deformed nuclei \cite{Gor2019}:

\begin{equation}
\label{EQ_10}
\begin{array}{l}
{E_{r} =E_{r,{\rm sys}}^{\alpha } =}\\
{\hspace{0.5 cm} =e_{1}^{\alpha } \cdot (1-I^{2} )^{1/2} \cdot A^{-1/3} /(1+e_{2}^{\alpha } A^{-1/3} )^{1/2}.}
\end{array}
\end{equation}

It was found from fitting (the energy in MeV): $ e_{1}^{\rm{SLO}} =130.0\pm 0.9$, $e_{2}^{\rm{SLO}} =9.0\pm 0.2$ for SLO model and $e_{1}^{\rm{SMLO}} =128.0\pm 0.9$,   $e_{2}^{\rm{SMLO}} =8.5\pm 0.2$ for SMLO model.

The expression (\ref{EQ_10}) is similar to that obtained as a good approximation to eigenenergy of dynamical equation for the GDR vibrations within liquid droplet hydrodynamic model \cite{Mye1977} (used in the TLO model, eq.(\ref{EQ_16})) and it also corresponds to a sum rule prescription for GDR energy \cite{Lip1988,Lip1989}.

The GDR energies  $E_{r,1} $,  $E_{r,2} $ are determined by the energies of the vibrations along and perpendicular  to the symmetry axis: $E_{r,1}=E_{a}$, $E_{r,2}=E_{b}$ (for prolate nuclei, i.e. $\beta _{2} > 0$ ) and $E_{r,1} =E_{b}$, $E_{r,2} =E_{a}$ (for oblate nuclei, i.e. $\beta_{2} < 0$). The systematics of the GDR resonance energies $E_{a}$ and $E_{b}$  are calculated by the use of the expressions \cite{Gor2019}

\begin{equation}
\label{EQ_11}
E_{a}= 3 E_{r} /(1+2 D),\ \ E_{b} =3 E_{r} D/(1+2 D),
\end{equation}

\noindent which are resulted from (\ref{EQ_8}), (\ref{EQ_9}) for known ratio $D=E_{b} /E_{a} $. Here  (and below in clear situations) index $\alpha $ denoting  PSF model is omitted.  The expression within hydrodynamical model \cite{Dan1958} is used  for $D$,

\begin{equation}
\label{EQ_12}
D=E_{b}/E_{a}=0.911\cdot R_{a}/R_{b}+0.089,
\end{equation}

\noindent  with $R_{a}$ ($R_{b}$) for a length of the ellipsoid semi-axis (semi-axes of speroid) along (perpendicular) rotational symmetry axis: $R_{a} /R_{b} =(1+\alpha _{2} )/(1-\alpha _{2} /2)$, $\alpha _{2} $ is the quadrupole deformation parameter $\alpha _{2}=\beta _{2}\sqrt{5/4\pi}$.   Notice that within the linear approximation on deformation and $D\approx R_{a}/R_{b} $, the general expressions  for resonance energies of normal modes of giant vibrations from (\ref{EQ_11})  coincide with expressions of the nuclear hydrodynamics model \cite{Eis1987}:

\begin{equation}
\label{EQ_13}
E_{a}=E_{r} \cdot \frac{R_{0} }{R_{a} },\ \ E_{b}=E_{r} \cdot \frac{R_{0} }{R_{b} } ,
\end{equation}

\noindent where $R_{0}$ is the radius of a sphere of equal volume $R_{0}^{3} =R_{a} \cdot R_{b}^{2} $.

For resonance width systematics,  the simplest power law expression was taken

\begin{equation}
\label{EQ_14}
\Gamma _{r,j}^{\alpha } =\Gamma _{r,j,{\rm sys}}^{\alpha } =c(\alpha )\; \cdot (E_{r,j} )^{d(\alpha )} \ \ {\rm  MeV},
\end{equation}

\noindent  with fitting results: $c({\rm SMLO})=0.42\pm 0.05$, $d({\rm SMLO})=0.90\pm 0.04$; and $c({\rm SLO})=0.32\pm 0.03$, $d({\rm SLO})=0.98 \pm 0.03$;  $c({\rm GLO})=c({\rm SLO})$; $d({\rm GLO})=d({\rm SLO})$.

Approximation of a triaxial ellipsoid is used in the TLO model for nuclear shape in deformed nuclei \cite{Jun2008,Gro2017}, and the nuclear shapes are determined in terms of $\beta $ and the triaxiality angle $\gamma $,  where $\beta $ represents the extent of quadrupole deformation (describes the deviation from sphere) and $\gamma $ gives the degree of axial asymmetry. In this approach, the GDR splits into three components and the E1 PSF is described by the expression (\ref{EQ_3}) with $j_{m} =3$. Similar general expression  was originally proposed in refs.\cite{Alh1988,Alh1990} for description of photoabsorption cross-sections of fast rotating nuclei in a triaxial nuclear shape approach.

For the input parameters of the TLO PSF,  the theoretical expressions were  taken from different theoretical models. For the resonance energies of the normal modes, the expressions (\ref{EQ_13}) of the nuclear hydrodynamics model were adopted\cite{Jun2008,Gro2017}:

\begin{equation}
\label{EQ_15}
E_{r,j}^{{\rm TLO}} =E_{r}^{{\rm LDH}} \frac{R_{0} }{R_{j} } ,
\end{equation}

\noindent  where $E_{r}^{{\rm LDH}}$ is the resonance energy  of the equivalent in volume spherical nucleus with the radius $R_{0}$. Droplet mode solution of liquid droplet hydrodynamic model \cite{Mye1977} was taken for energy $E_{r}^{{\rm LDH}}$,

\begin{equation}
\label{EQ_16}
\begin{array}{l}
{\displaystyle E_{r}^{{\rm LDH}} =\frac{\hbar c}{R_{0} } \sqrt{\frac{8J}{m^{*} c^{2} } \frac{1}{(1-I^{2} )} } \times} \\
{\displaystyle\hspace{0.3 cm} \times \left[{1+u-\varepsilon \frac{1+\varepsilon +3u}{1+\varepsilon +u} }\right]^{-1/2} \hspace{0.5 cm} {\rm MeV},}
\end{array}
\end{equation}

\noindent with $R_{0} =1.16 \cdot A^{1/3}$, $r_{0}=1.16$ fm, $\varepsilon =0.0768$, $u=(1-\varepsilon )A^{-1/3} 3J/Q$, symmetry-energy value $J=32.7$ MeV and effective surface-stiffness value $Q=29.2$ MeV; $m^{*} c^{2} =874$ MeV. General expression of  simplified version of the eq.(\ref{EQ_16}) (see \cite{Mye1977}, Eq.(5.1)) is similar to the expression (\ref{EQ_10}) but with different parameters.

\begin{figure}[htbp]
    \includegraphics[width=0.95\columnwidth]{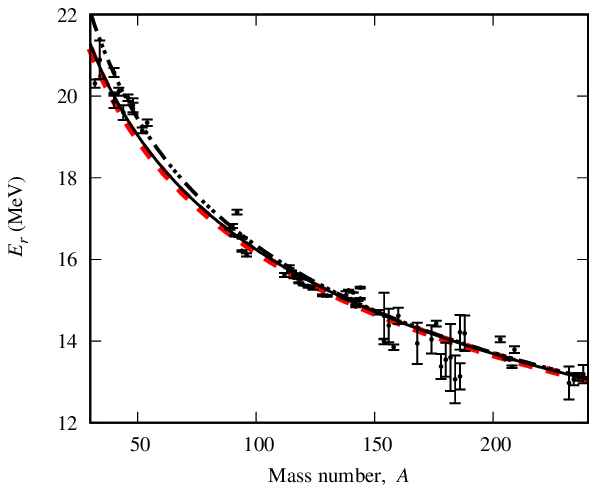}
    \includegraphics[width=0.95\columnwidth]{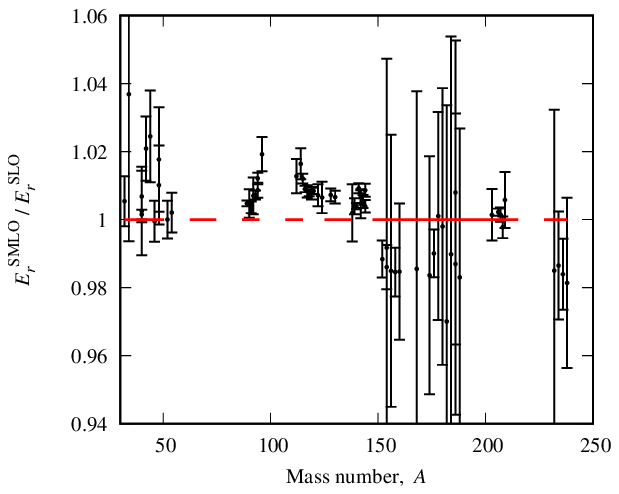}
	\caption{Comparison of the experimental values of average GDR energies ($E_{r}^{{\rm SLO(SMLO)}} $) \cite{Plu2018} with calculations within systematics (\ref{EQ_10}) for SLO and SMLO models ($E_{r,{\rm sys}}^{{\rm SLO(SMLO)}} $) and  within liquid droplet hydrodynamic model ($E_{r}^{{\rm LDH}} $). Upper panel:  $E_{r}^{{\rm SLO}} $ - black dots with errorbars; $E_{r,{\rm sys}}^{{\rm SLO}} $ - black solid line ($\bm{^{\_ \_ \_ \_ \_ \_ \_ } }$)  , $E_{r,{\rm sys}}^{{\rm SMLO}} $- red dashed line ($\bm{{\color{red} ----}}$)  , $E_{r}^{{\rm LDH}} $ - black dashed followed by three dots line ($\bm{-\, \cdot \, \cdot \, \cdot -\, \cdot \, \cdot \cdot \, -}$). Bottom panel shows the ratios of average experimental energies within SMLO and SLO models.}
	\label{fig_2}
\end{figure}

Figure~\ref{fig_2} shows the experimental values of the average GDR energies (\ref{EQ_8}) with uncertainties for SLO and SMLO models as a function of mass number in spherical nuclei and deformed nuclei with 150$<A<$190, 220$<A<$253. The data are compared with calculations within  systematics (\ref{EQ_10}), $E_{r,{\rm sys}}^{{\rm SLO(SMLO)}}$, and within  liquid droplet hydrodynamic model (\ref{EQ_16}), $E_{r}^{{\rm LDH}}$. In the calculations the neutron-proton asymmetry factor was taken as corresponding to Green's approximation \cite{Gre1953} to the line of $\beta $ stability:
$I=(N-Z)/A=0.4A/(A+200)$.

One can see that experimental average GDR energies determined by the use of SLO and SMLO approaches are in rather close agreement and their relative deviation does not exceed $\sim$6\%. Relative deviation of systematical values  of these average energies is less then 1\%. Droplet mode solution (\ref{EQ_16}) for resonance energies  good reproduce systematical values of experimental data for nuclei with $A>80$.

The Hill-Wheeler parameterization \cite{Has1988}  is used for calculation of the semi-axis lengths $R_{j}$ in (\ref{EQ_15}), in \cite{Jun2008}

\begin{equation}
\label{EQ_17}
R_{j} =R_{j}^{{\rm H}} =R_{0} \cdot \exp \left({\sqrt{5/4\pi } \cdot \beta \cdot \cos (\gamma -\frac{2}{3} j \pi )}\right).
\end{equation}

The radius parameters $R_{j}$, which are inversely proportional to the harmonic oscillator constants, were used in ref.\cite{Gro2017}, but it was indicated in this references that the differences in the results using (\ref{EQ_17}) were not significant.

The width $\Gamma _{j}^{{\rm TLO}} $ was independent of gamma-ray energy and the expression with a power law dependence on the resonance energy was used \cite{Jun2008,Gro2017,Jun2010}:

\begin{equation}
\label{EQ_18}
\Gamma _{j}^{{\rm TLO}} =0.045(E_{r,j}^{{\rm TLO}} )^{1.6} \ \ {\rm MeV}.
\end{equation}

\begin{figure}[htbp]
    \includegraphics[width=0.95\columnwidth]{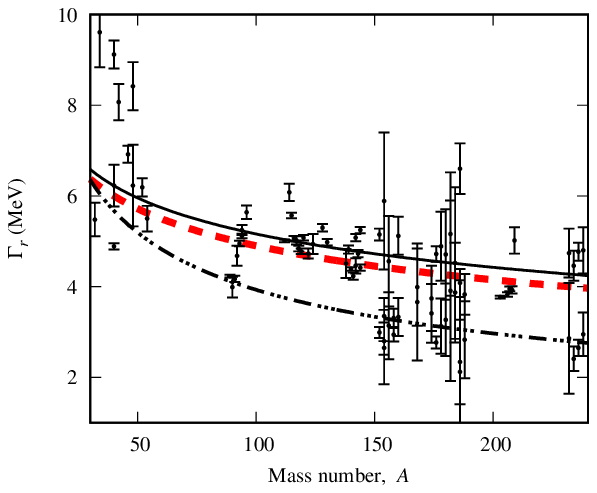}
    \includegraphics[width=0.95\columnwidth]{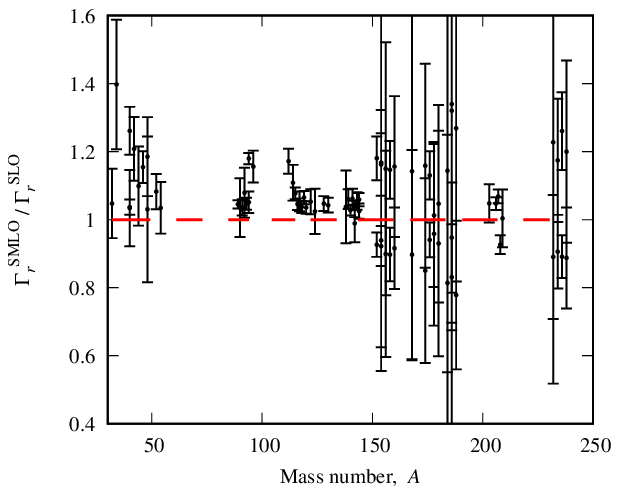}
	\caption{Comparison of the experimental values of the GDR widths ($\Gamma _{r}^{{\rm SLO(SMLO)}} $) \cite{Plu2018} with uncertainties with calculations within systematics (\ref{EQ_14}) for SLO and SMLO models ($\Gamma _{r,{\rm sys}}^{{\rm SLO(SMLO)}} $) and  within TLO model ($\Gamma _{r}^{{\rm TLO}} $). Upper panel: experimental GDR widths $\Gamma _{r}^{{\rm SLO}} $ - black dots with errorbars, $\Gamma _{r,{\rm sys}}^{{\rm SLO}} $ - black solid line ($\bm{^{\_ \_ \_ \_ \_ \_ \_ } }$),  $\Gamma _{r,{\rm sys}}^{{\rm SMLO}} $ - red dashed line ($\bm{{\color{red} ----}}$),  $\Gamma _{r}^{{\rm TLO}} $ is black dashed followed by three dots ($\bm{-\, \cdot \, \cdot \, \cdot -\, \cdot \, \cdot \cdot \, -}$) $\Gamma ^{{\rm TLO}} $. Bottom panel: the ratios of average experimental widths \cite{Plu2018} within SMLO and SLO models.}
	\label{fig_3}
\end{figure}

Figure \ref{fig_3} demonstrates the experimental values of the GDR energies ($\Gamma _{r}^{{\rm SLO(SMLO)}} $) \cite{Plu2018} with uncertainties for SLO and SMLO models as a function of mass number in spherical nuclei and in deformed nuclei with 150$<A<$190, 220$<A<$253. The data are compared with calculations within  systematics for SLO and SMLO models (\ref{EQ_14}), $\Gamma _{r,{\rm sys}}^{{\rm SLO(SMLO)}}$, and within  TLO model (\ref{EQ_19}), $\Gamma_{r}^{{\rm TLO}}$. In the calculations the neutron-proton asymmetry factor was taken as corresponding to Green's approximation \cite{Gre1953} to the line of $\beta$ stability.

One can see that experimental  GDR widths determined by the use of SLO and SMLO approaches are generally in agreement with allowance for uncertainties but their relative deviation can reach $\sim$40\%. Systematical values  of the SLO and SMLO model describe experimental data better then TLO approach. Systematics and TLO expressions not very good reproduce experimental data in isotopes with mass-numbers  $A\le 50$.

The approximation of equally probable excitation of normal modes of the giant collective vibration were taken  for the weights:
$s_{j}^{{\rm TLO}} =s^{{\rm TLO}} /3$ with $s^{{\rm TLO}} =1.02$ for the sum of the weights.

\section{Particularities and modifications of the PSF models  with energy-dependent width for keeping energy-weighted sum rule}
\label{sec:2}

Intensive studies of the photoabsorption in middle-weight to heavy nuclei demonstrated that the photoabsorption cross-sections at the low-energy tail of the GDR can be better described with allowance for increasing dependence of the width $\Gamma (\varepsilon _{\gamma } )$ on gamma-ray energy (see \cite{RIPL,RIPL2,RIPL3,Plu2018} for references). However, if the width  increases with energy steadily, the total integral of the line-shape functions $F_{j}^{\alpha } $, (\ref{EQ_5}), over energy to infinity can not be equal to unity, which is needed for fulfilment of the energy-weighted sum rule (EWSR) for electric dipole gamma-transitions.

The EWSR ($S_{{\rm EWSR}}$) constraints the energy integrated total photoabsorption cross-section $\sigma_{{\rm E1}}(\varepsilon_{\gamma})$,

\begin{equation}
\label{EQ_19}
\sigma _{{\rm int}} =\int _{0}^{\infty }\sigma _{{\rm E1}} (\varepsilon _{\gamma } ) d\varepsilon _{\gamma } =\frac{8\pi \alpha }{3} S_{{\rm EWSR}},
\end{equation}

\noindent where $\alpha =e^{2} /(\hbar c)$ is the fine structure constant. The energy integrated cross-section can be presented in the form

\begin{equation}
\label{EQ_20}
\sigma _{{\rm int}} \, =\sigma _{{\rm TRK}} \cdot s,
\end{equation}

\noindent where a factor $s$ determines deviation of $\sigma _{{\rm int}} $ on TRK sum rule (\ref{EQ_4}). It was mentioned above, that in nonrelativistic potential interaction approach with absence of the velocity-dependent and exchange forces $s=1$, but $s \sim 1.2$  in the presence of he velocity-dependent and exchange forces.

One can see from eqs.(\ref{EQ_2}-\ref{EQ_5}), that

\begin{equation}
\label{EQ_21}
\begin{array}{c}
{\displaystyle \sigma _{{\rm int}} =\int _{0}^{\infty }\sigma _{{\rm E1}} (\varepsilon _{\gamma } ) d\varepsilon _{\gamma } =} \\
{\displaystyle \hspace{0.3 cm} =\sum _{j=1}^{j_{m} }\sigma _{{\rm TRK}} s_{j}^{\alpha } I_{j}^{\alpha } (\varepsilon _{\gamma } \rightarrow \infty ),} \\
{\displaystyle I_{j}^{\alpha } (\varepsilon _{{\rm max} } )=\int _{0}^{\varepsilon _{{\rm max} } }F_{j}^{\alpha } (\varepsilon _{\gamma } ) \, d\varepsilon _{\gamma},}
\end{array}
\end{equation}

\noindent and the EWSR is fulfilled for theoretical models $\alpha $ with the Lorentzian shape  if the integral $I_{j}^{\alpha}(\varepsilon_{\max})$   tends to unity at $\varepsilon _{{\rm max }} \rightarrow \infty$.

\begin{figure}[htbp]
    \includegraphics[width=0.95\columnwidth]{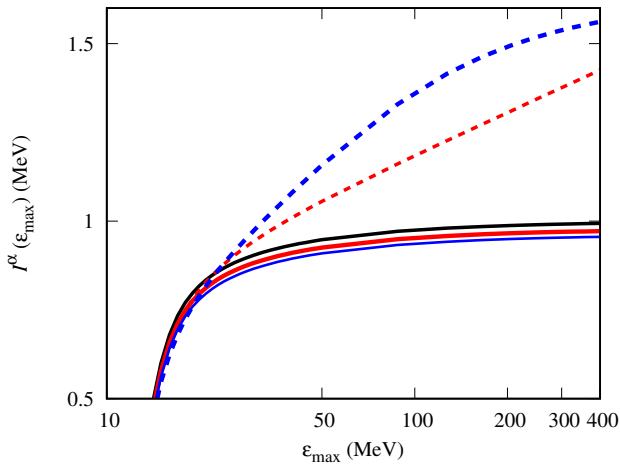}
	\caption{The value of integrals $I^{\alpha } $ calculated within the following models for GDR contribution to the PSF: $\bm{^{\_ \_ \_ \_ \_ \_ \_ } }$ (black solid line) SLO,  $\bm{{\color{red} ----}}$  (red dashed line)   SMLO,  $\bm{{\color{red}^{\_ \_ \_ \_ \_ \_ \_ } }}$  (solid red line)  SMLOe,  $\bm{{\color{blue} ----}}$ ( blue dashed  line)  GLO,  $\bm{{\color{blue}^{\_ \_ \_ \_ \_ \_ \_ } }}$ ( blue solid line) GLOe.}
	\label{fig_4}
\end{figure}

\begin{figure}[htbp]
    \includegraphics[width=0.95\columnwidth]{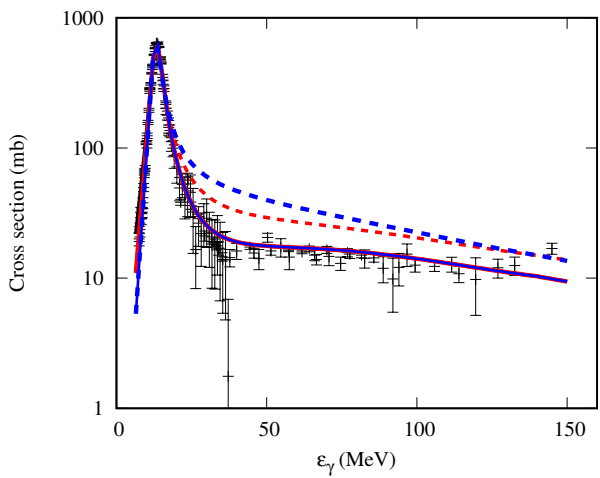}
	\caption{Photoabsorption cross-sections for lead as a function of the gamma-ray energy. Theoretical calculations are performed with the GDR component within the models: $\bm{^{\_ \_ \_ \_ \_ \_ \_ } }$ (black solid line) SLO,  $\bm{{\color{red}----}}$  (red dashed line)   SMLO,  $\bm{{\color{red}^{\_ \_ \_ \_ \_ \_ \_ } }}$  (red solid line)  SMLOe,  $\bm{{\color{blue}----}}$ ( blue dashed  line)  GLO,  $\bm{{\color{blue}^{\_ \_ \_ \_ \_ \_ \_ } }}$ ( blue solid line) GLOe.  The experimental data are taken from refs.\cite{Ahr1985,Ish2013}.}
	\label{fig_5}
\end{figure}

The figures~\ref{fig_4},\ref{fig_5} show the values of the integrals $I^{\alpha } (\varepsilon _{{\rm max}})$ as a function of the energy $\varepsilon _{{\rm max}}$ and photoabsorption cross-sections in dependence of gamma-ray energy in nucleus ${}^{208}$Pb ($j_{m}=1$). The photoabsorption cross-sections were calculated for different models  for  GDR component and quasi-deuteron photodisintegration was taken into account. The GDR parameters were taken from the first dataset for ${}^{208}$Pb in the table~1 of the ref.~\cite{Plu2018}. In the models of SMLOe and GLOe the general expressions of the approaches of SMLO and GLO are used with modificated expressions for shape widths above the GDR energy (see below eq.(\ref{EQ_22})).

It can be seen from figs.~\ref{fig_4},\ref{fig_5} that, if $\Gamma (\varepsilon _{\gamma } )$ is permanently increasing, the integrals $I^{\alpha } (\varepsilon _{\max } )$ of the Lorentzian-like function over energy with energy dependent width are very close to SLO values for gamma-ray till $\sim$30 MeV but  for the gamma-ray energies  $\gtrsim$30 MeV they can be more than unity due to overestimation of the GDR component of the photoabsorption cross-section \cite{Gro2017}. It  leads to violation of the EWSR.

Note that, the expressions for energy-dependent width which are increased with energy are based on a low-energy approximation of nucleon-nucleon collision cross-section in nuclear medium and should be corrected at high energies. A behavior of the damping width at high energies can be analyzed under the assumption that it corresponds to the width for damping 1 particle-1 hole states into 2 particle-2 hole states \cite{Dan1965,Gar1984,Kol1996,Plu2001,Kal1986}. The latter width is proportional to product of the mean square residual matrix element ($M^{2}$) and the density of the 2p-2h states which is proportional to the square of gamma-ray energy with the results that $\Gamma (\varepsilon _{\gamma } )\propto M^{2} (\varepsilon _{\gamma } )\cdot \varepsilon _{\gamma }^{2} $. The average squared matrix element was studied within the exciton model \cite{Kon2004,Del2010}. It was shown that $M^{2} $ is energy constant at low energies and depends inversely on cube of energy at high energies. It means that the damping width increases initially with squared energy and then decreases in inverse proportion to the energy at high energies. Here, we simulate this effect  for  width $\Gamma (\varepsilon _{\gamma } )$ of the Lorentzian shapes of the SMLO and GLO models in very simple way and use constant width after GDR energies $\varepsilon _{\gamma} > E_{r,j}$:

\begin{equation}
\label{EQ_22}
\Gamma _{j}^{\bar{\alpha }} (\varepsilon _{\gamma } )=\left\{\begin{array}{l} {\Gamma _{j}^{\alpha } (\varepsilon _{\gamma } ){\kern 1pt} {\kern 1pt} ,{\kern 1pt} \, \, \, \, \, \, \, \, \, \, \, \, \, \, \, \, \, \, \, \, \, \, \, {\kern 1pt} \varepsilon _{\gamma } \le E_{r,j}^{\alpha } ,} \\ {\Gamma _{j}^{\alpha } (E_{r,j}^{\alpha } )=\Gamma _{r,j}^{\alpha } {\kern 1pt} {\kern 1pt} ,{\kern 1pt} {\kern 1pt} {\kern 1pt} \varepsilon _{\gamma } >E_{r,j}^{\alpha } } \end{array}\right. .
\end{equation}

Here these models are denoted  as extended SMLO and GLO models (SMLOe and GLOe) and in eq.~(\ref{EQ_22}) $\bar{\alpha }$=SMLOe (GLOe) if $\alpha $=SMLO(GLO) . One can see from the figs.~\ref{fig_4},\ref{fig_5} that the models with energy-dependent confined widths (\ref{EQ_22}) can correct in a very simple way an inaccurate high-energy behavior of the GDR component of the photonuclear cross-sections. They also can lead in good approximation to correct value of the EWSR. In the energy range $\varepsilon _{\gamma } <30$ MeV the calculations within SMLOe (GLOe) model agree with the results for SMLO (GLO) model. So, the approximation of the energy-dependent width restricted to GDR value can be recommended for using in the nuclear reaction codes for more correct modeling of the E1 photon strength function at high energy tail of the GDR ($\varepsilon _{\gamma } \gtrsim 30$ MeV).

\section{Calculations and results}
\label{sec:3}

Quantitative comparison  between  photoabsorption experimental data with calculations by different Lorentzian --type  PSF models are given below for the 88 even-even isotopes from ${}^{24}$Mg till ${}^{238}$U listed in the table~\ref{tabl1} of the  ref.~\cite{Plu2018} and used for determination of the recommended GDR parameters. For these nuclei, experimental data on total photoabsorption cross-sections $\sigma (\gamma,{\rm abs})$ from \mbox{EXFOR} are taken and references are also indicated in the table~1 of \cite{Plu2018}.

Analytical PSF expressions for models of SLO, GLO, SMLO, SMLOe and their input parameters  were described in previous section. Three sets of the deformation parameters are used for TLO model, and corresponding calculation results are denoted as TLO(1), TLO(2) and TLO(3). The deformation parameters for TLO(1) and TLO(2) models were taken from the data-files of ref.\cite{Jun2010}. The  TLO(3) was used for PSF calculations for isotopes specified in the table~\ref{tabl1} with indicated deformation parameters.

\begin{table*}[t]
\center
\caption{{The values of deformation parameters used for PSF calculations for listed nuclei.}}
\label{tabl1}
\begin{tabular}[c]{ccccccccc}
\hline\noalign{\smallskip}
Nuclei & \multicolumn{1}{p{1.4in}}{ SLO, SMLO, GLO: \newline $\beta _{2}$  \cite{RIPL2,RIPL3} }& \multicolumn{2}{p{1.4in}}{ TLO(1):\newline $\beta $=$\beta _{{\rm H}}^{{\rm B}}$,
$\gamma=\gamma _{{\rm H}}^{{\rm B}}$ \cite{Jun2010}; \newline HFB \newline } & \multicolumn{2}{p{1.4in}}{TLO(2):\newline $\beta $=$\beta _{{\rm C}}^{{\rm B}} $, $\gamma $=$\gamma _{{\rm C}}^{{\rm B}} $ \cite{Jun2010};\newline  CHFB+5DCH } & \multicolumn{3}{p{1.6in}}{TLO(3):\newline $\beta $=$\beta^{{\rm H}} $, $\gamma $=$\gamma ^{{\rm H}}$ fitted \newline parameters taken from \newline indicated refs.} \\
\noalign{\smallskip}\hline\noalign{\smallskip}
                 & $\beta _{2} $ & $\beta $ & $\gamma $ & $\beta $ & $\gamma $ & $\beta $ & $\gamma $ & ref. \\
${}^{94}$Mo & 0.053 & 0.0 & 0$^\circ$ & 0.16 & 28$^\circ$ & -0.08 & 20$^\circ$ & \cite{Jun2008} \\
${}^{98}$Mo & 0.184 & 0.0 & 0$^\circ$ & 0.208 & 26$^\circ$ & 0.18 & 23$^\circ$ & \cite{Jun2008} \\
${}^{146}$Nd & 0.173 & 0.165 & 0$^\circ$ & 0.167 & 25$^\circ$ & 0.17 & 26$^\circ$ & \cite{Gro2009} \\
${}^{148}$Sm & 0.171 & 0.167 & 0$^\circ$ & 0.169 & 25$^\circ$ & 0.13 & 25$^\circ$ & \cite{Gro2010} \\
${}^{156}$Gd & 0.295 & 0.343 & 0$^\circ$ & 0.347 & 10$^\circ$ & 0.22 & 11$^\circ$ & \cite{Gil1965} \\
${}^{168}$Er & 0.292 & 0.346 & 0$^\circ$ & 0.361 & 9$^\circ$ & 0.28 & 12$^\circ$ & \cite{Gro2010} \\
${}^{190}$Os & 0.153 & 0.175 & 33$^\circ$ & 0.188 & 25$^\circ$ & 0.16 & 21$^\circ$ & \cite{Gro2010} \\
${}^{196}$Pt & -0.135 & 0.13 & 54$^\circ$ & 0.135 & 32$^\circ$ & 0.13 & 29$^\circ$ & \cite{Jun2008} \\
${}^{206}$Pb & -0.008 & 0.0 & 0$^\circ$ & 0.058 & 25$^\circ$ & 0.02 & 40$^\circ$ & \cite{Gil1965} \\
${}^{238}$U & 0.236 & 0.272 & 0$^\circ$ & 0.292 & 8$^\circ$ & 0.29 & 17$^\circ$ & \cite{Gro2009} \\
\noalign{\smallskip}\hline
\end{tabular}
\end{table*}

Table~\ref{tabl1} shows also the deformation parameters used for these isotopes as input for TLO(1) and TLO(2) models. They were calculated for even-even nuclides within framework of the  Hartree-Fock-Bogoliubov theory (HFB; TLO(1) - $\beta $=$\beta _{{\rm H}}^{{\rm B}} $,$\gamma $=$\gamma _{{\rm H}}^{{\rm B}} $) and Constrained Hartree-Fock-Bogoliubov approach with five-dimensional collective Hamiltonian (CHFB+5DCH; TLO(2)- $\beta=\beta _{{\rm C}}^{{\rm B}}$, $\gamma=\gamma _{{\rm C}}^{{\rm B}}$). The Bohr parameterization \cite{Has1988} of the axis lengths was adopted in these calculations of the  deformation parameters

\begin{equation}
\label{EQ_23}
R_{j} =R_{j}^{{\rm B}} =R_{0} \left({1+\sqrt{5/4\pi } \cdot \beta \cos (\gamma -\frac{2}{3}  j \pi)}\right).
\end{equation}

Therefore  we used this expression for semiaxes in TLO(1) and TLO(2) models.  In first order on deformation parameters the values of   the semiaxes  (\ref{EQ_17}), (\ref{EQ_23}) are in agreement.

Two criteria were taken for comparison of a quality of the description of the experimental photonuclear data using different E1 PSF models:
1) minimum of the least-square deviation $\chi_{\alpha}^{2}$ , and
2) minimum of the root-mean-square (rms) deviation factor  $f_{\alpha}$ \cite{Gor2018,Bei1974}:

\begin{equation}
\label{EQ_24}
\chi _{\alpha }^{2} =\frac{1}{n} \sum _{i=1}^{n}\frac{(\sigma _{\rm exp } (\varepsilon _{i} )-\sigma _{{\rm the}}^{\alpha } (\varepsilon _{i} ))^{2} }{(\Delta \sigma (\varepsilon _{i} ))^{2}},
\end{equation}

\begin{equation}
\label{EQ_25}
\begin{array}{c}
{\displaystyle f_{\alpha } =\exp \{ \chi _{\ln ,\alpha } \} ,} \\
 {\displaystyle \chi _{\ln ,\alpha }^{2} =
 \frac{1}{n} \sum _{i=1}^{n}\{ \ln \sigma _{{\rm the}}^{\alpha } (\varepsilon _{i} )-\ln \sigma _{{\rm exp}} (\varepsilon _{i} ) \} ^{2} = }\\
{\displaystyle \hspace{0.3 cm}  =\frac{1}{n} \sum _{i=1}^{n}\ln ^{2} (\frac{\sigma _{{\rm the}}^{\alpha } (\varepsilon_{i})}{\sigma _{{\rm exp}} (\varepsilon _{i})}).}
 \end{array}
\end{equation}

Here $\sigma_{{\rm the}}^{\alpha}(\varepsilon_{i})=\sigma_{{\rm GDR}}^{\alpha} (\varepsilon_{i})+\sigma_{{\rm qd}}(\varepsilon_{i})$ is the theoretical cross section, eqs.~(\ref{EQ_1})-(\ref{EQ_3}), at $\gamma$-ray energy $\varepsilon_{i}$;
$\sigma_{{\rm exp}} (\varepsilon_{i})$ is the experimental photoabsorption cross section from the EXFOR database; $\Delta\sigma(\varepsilon_{i} )$ is the  data uncertainty; $n$ is the number of experimental data points.

The rms deviation factor $f_{\alpha}$ corresponds to criterion of a minimum at given gamma-ray energy of a weighted sum of squared deviations of the theoretical cross-section from their experimental data in natural logarithmic scale. Because of different estimations of uncertainties in different datasets and, as a rule, for lack of reliable estimations of the systematic errors, a weight of every point is taken as empirical probability $1/n$. A logarithmic scale is used due to large-range changes of the photoabsorption cross-sections.

The $\chi_{\alpha}^{2}$ and $f_{\alpha}$ values were calculated in the gamma-ray energy intervals from 5 MeV ( or minimal value of the energy $>$5 MeV) till 30 MeV (or maximal value of energy $<$30 MeV).

If the EXFOR datafiles were not contained experimental uncertainties, the energy-dependent relative uncertainties $\delta\sigma(\varepsilon_{i})=\Delta\sigma(\varepsilon_{i})/\sigma_{\rm exp}(\varepsilon_{i})$ were taken in accordance with ref.~\cite{Plu2018}. The energy dependence was chosen to simulate the statistical error that is inversely proportional to the counting rate which is maximum near the GDR. Hence, the energy-dependent relative uncertainties were assumed to take minimum values (10\%) near the GDR peaks and maximum values (50\%) on the GDR tails. For spherical nuclei, a triangular dependence on gamma-energy was assumed, while for deformed nuclei a trapezoidal dependence with the GDR peaks as the top corners of the trapezium.

\begin{figure}[htbp]
    \includegraphics[width=0.95\columnwidth]{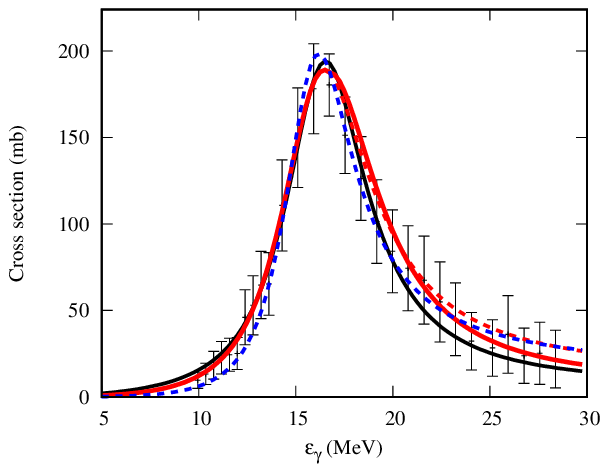}
    \includegraphics[width=0.95\columnwidth]{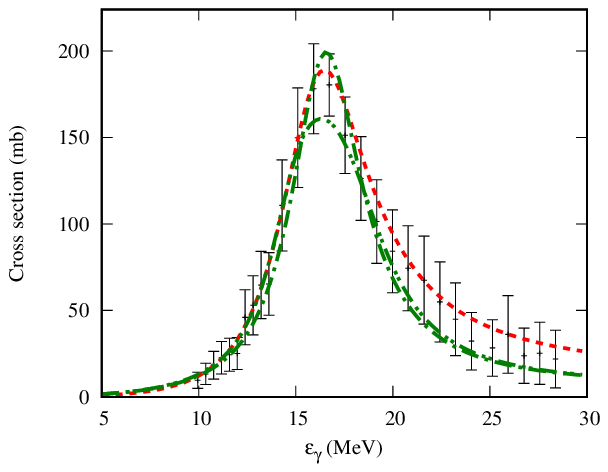}
	\caption{The experimental cross section data for ${}^{94}$Mo in comparison with calculations within different PSF models. Upper figure: $\bm{^{\_ \_ \_ \_ \_ \_ \_ } }$ (black solid line) SLO,  $\bm{{\color{red}----}}$  (red dashed line)   SMLO,  $\bm{{\color{red}^{\_ \_ \_ \_ \_ \_ \_ }} }$  (solid red line)  SMLOe,  $\bm{{\color{blue}----}}$ ( blue dashed  line)  GLO. Bottom figure: $\bm{{\color{red}----}}$  (red dashed line)   SMLO;  $\bm{{\color{green}-\, \cdot \, -\, \cdot \, -}}$ (green dashed followed by one dot line) TLO(1); $\bm{{\color{green}-\, \cdot \, \cdot \, -\, \cdot \, \cdot \, -}}$ (green dashed followed by two dots line) TLO(2);  $\bm{{\color{green}-\, \cdot \, \cdot \, \cdot -\, \cdot \, \cdot \cdot \, -}}$ (green dashed followed by three dots line) TLO(3). Experimental data are taken from ref.\cite{Gur1981}.}
	\label{fig_6}
\end{figure}

\begin{figure}[htbp]
    \includegraphics[width=0.95\columnwidth]{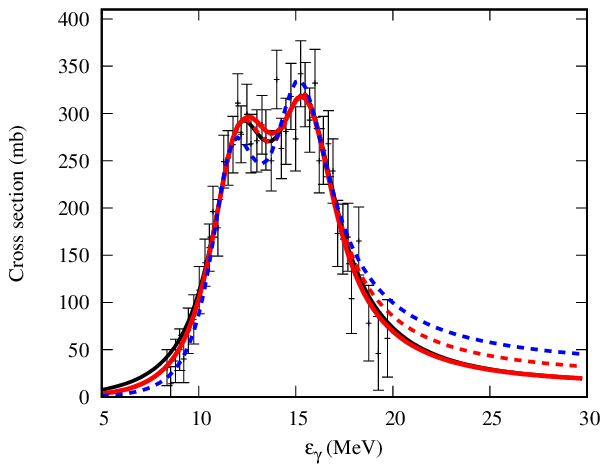}
    \includegraphics[width=0.95\columnwidth]{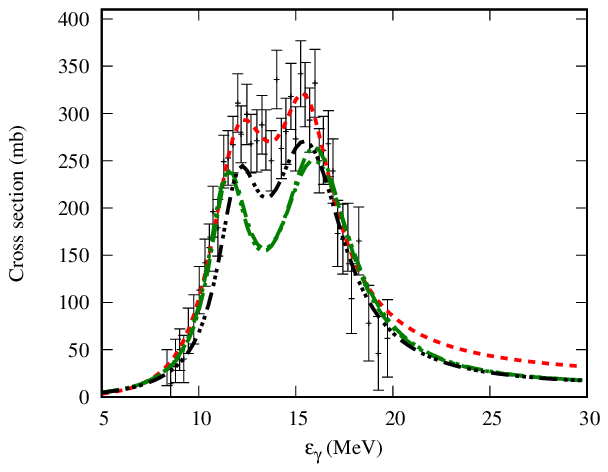}
	\caption{The experimental cross section data for ${}^{168}$Er in comparison with calculations within different PSF models. Upper figure: $\bm{^{\_ \_ \_ \_ \_ \_ \_ } }$ (black solid line) SLO,  $\bm{{\color{red} ----}}$  (red dashed line)   SMLO,  $\bm{{\color{red}^{\_ \_ \_ \_ \_ \_ \_ } }}$  (solid red line)  SMLOe,  $\bm{{\color{blue}----}}$ ( blue dashed  line)  GLO. Bottom figure: $\bm{{\color{red} ----}}$  (red dashed line)   SMLO;  $\bm{{\color{green} -\, \cdot \, -\, \cdot \, -}}$ (green dashed followed by one dot line) TLO(1); $\bm{{\color{green} -\, \cdot \, \cdot \, -\, \cdot \, \cdot \, -}}$ (green dashed followed by two dots line) TLO(2);  $\bm{{\color{green} -\, \cdot \, \cdot \, \cdot -\, \cdot \, \cdot \cdot \, -}}$ (green dashed followed by three dots line) TLO(3). Experimental data are taken from ref.\cite{Kad1983}.}
	\label{fig_7}
\end{figure}

\begin{table*}[htbp]
\center
\caption{{The values of the ratios  $\chi _{\alpha }^{2} /\chi _{{\rm SLO}}^{2} $ and $f_{\alpha } /f_{{\rm SLO}} $  for ${}^{94}$Mo and ${}^{168}$Er.}}
\label{tabl2}
\begin{tabular}[c]{ccccccccc}
\hline\noalign{\smallskip}
{Isotope} & {Criteria} & {Gamma-ray energy intervals} & {SMLO} & {SMLOe} & {GLO} & {TLO(1)} & {TLO(2)} & {TLO(3)} \\
\noalign{\smallskip}\hline\noalign{\smallskip}
{${}^{94}$Mo} & {$\chi _{\alpha }^{2} /\chi _{{\rm SLO}}^{2} $} & 5-30 MeV & 0.88 & 1.09 & 1.27 & 2.13 & 1.58 & 1.52 \\
{} &  & Near GDR peaks: 9.6 - 18.9 MeV & 0.78 & 1.27 & 1.74 & 1.78 & 1.46 & 1.00 \\
{} & {$f_{\alpha } /f_{{\rm SLO}} $} & 5-30 MeV & 0.95 & 0.91 & 1.06 & 1.06 & 1.02 & 1.05 \\
{} &  & Near GDR peaks: 9.6 - 18.9 MeV & 0.92 & 0.93 & 1.11 & 0.95 & 0.94 & 0.94 \\ {${}^{168}$Er} & {$\chi _{\alpha }^{2} /\chi _{{\rm SLO}}^{2} $} & 5-30 MeV & 0.95 & 0.89 & 1.36 & 6.93 & 7.49 & 4.22 \\
{} &  & Near GDR peaks: 10.9 - 18.8 MeV & 1.04 & 1.04 & 1.47 & 10.20 & 11.16 & 4.80 \\
{} & {$f_{\alpha } /f_{{\rm SLO}} $} & 5-30 MeV & 0.98 & 0.95 & 0.98 & 1.07 & 1.08 & 1.09 \\
{} &  & Near GDR peaks: 10.9 - 18.8 MeV & 1.01 & 1.00 & 1.04 & 1.23 & 1.25 & 1.10 \\ \noalign{\smallskip}\hline
\end{tabular}
\end{table*}

\begin{figure}[htbp]
    \includegraphics[width=0.95\columnwidth]{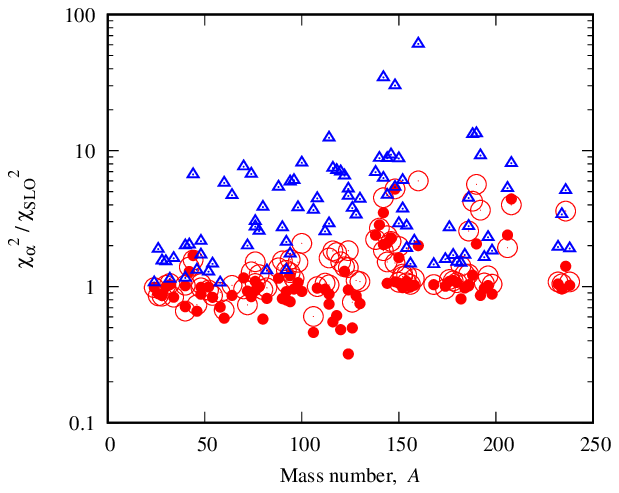}
    \includegraphics[width=0.95\columnwidth]{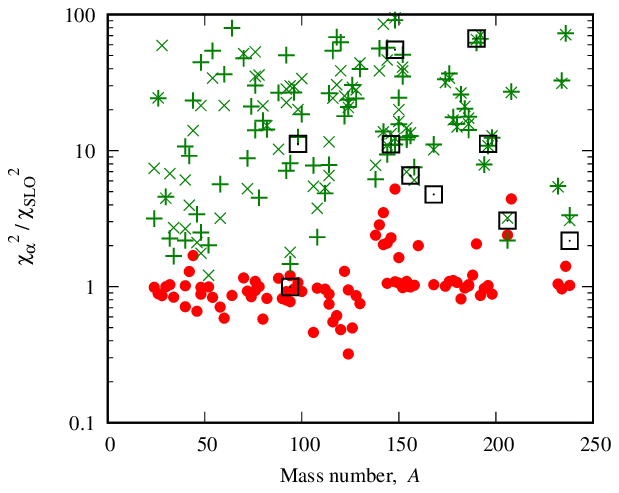}
	\caption{The relative least square values $\chi _{\alpha }^{2} /\chi _{{\rm SLO}}^{2} $ for even-even isotopes corresponding to calculations within different PSF models. Upper figure:  red circles ($\bm{{\color{red}\bullet}}$) are  SMLO; red empty  circles ($\bm{{\color{red}\odot }}$)  -  SMLOe;  blue empty triangles ($\bm{{\color{blue}\bigtriangleup }}$) -  GLO. Bottom figure: red circles ($\bm{{\color{red}\bullet}}$) -  SMLO; green crosses ($\bm{{\color{green}\times }}$) - TLO(1); green pluses (\textcolor{green}{\textbf{+}}) - TLO(2); black empty squares ($\bm{\boxdot }$) -  TLO(3). The presented results correspond to intervals from 5 MeV ( or minimal value of the energy $>$5 MeV) till 30 MeV (or maximal value of energy $<$ 30 MeV).}
	\label{fig_8}
\end{figure}

Figures~\ref{fig_6},\ref{fig_7} present the theoretical photoabsorption cross-sections obtained within different models of
$\sigma_{{\rm GDR}}^{\alpha}$ in comparison with experimental data $\sigma_{{\rm exp }} $ for the isotopes of ${}^{94}$Mo  and ${}^{168}$Er.
For these isotopes the least-square values and the $f_{\alpha}$ rms deviation factors for different models are given in the table~\ref{tabl2}.

\begin{figure}[htbp]
    \includegraphics[width=0.95\columnwidth]{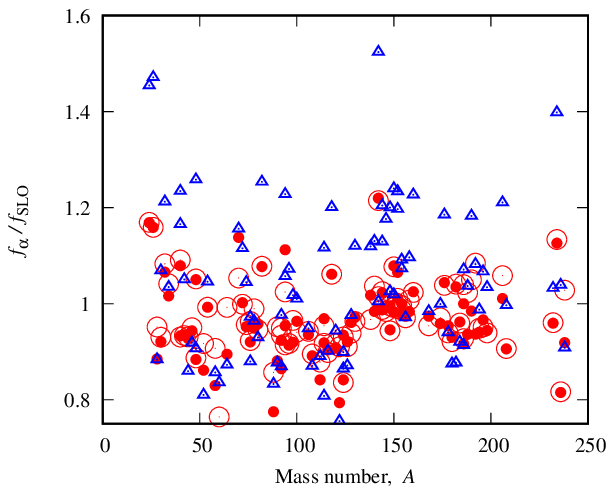}
    \includegraphics[width=0.95\columnwidth]{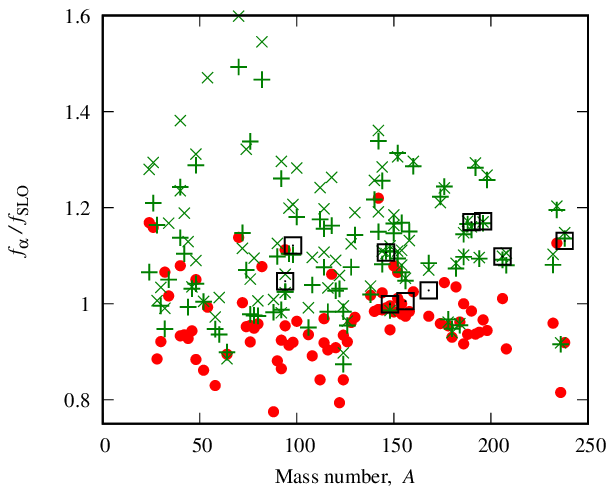}
	\caption{The relative values of rms deviation $f_{\alpha } /f_{{\rm SLO}} $ for even-even isotopes corresponding to calculations within different PSF models. Upper figure:  red circles ($\bm{{\color{red}\bullet}}$) are  SMLO; red empty  circles ($\bm{{\color{red}\odot }}$)  -  SMLOe;  blue empty triangles ($\bm{{\color{blue}\bigtriangleup }}$) -  GLO. Bottom figure: red circles ($\bm{{\color{red}\bullet}}$) -  SMLO; green crosses ($\bm{{\color{green}\times }}$) - TLO(1); green pluses (\textcolor{green}{\textbf{+}}) - TLO(2); black empty squares ($\bm{\boxdot }$) -  TLO(3).  The presented results correspond to intervals from 5 MeV ( or minimal value of the energy $>$5 MeV) till 30 MeV (or maximal value of energy $<$ 30 MeV).}
\label{fig_9}
\end{figure}

The relative least-square values and rms deviation factors obtained at comparisons of experimental data with theoretical calculations  are presented in figs.\ref{fig_8},\ref{fig_9}.
Table~\ref{tabl3} gives  arithmetic mean values $R_{\alpha } =<\chi _{\alpha }^{2} >/<\chi _{{\rm SLO}}^{2} >$ of least-square deviations for PSF model $\alpha $ to that for SLO model. The ratios $F_{\alpha } =<f_{\alpha } >/<f_{{\rm SLO}} >$ of arithmetic mean values of rms deviation factors $<f_{\alpha}>$  for  PSF model $\alpha $ to that for SLO model  are presented in the table~\ref{tabl4}.
The TLO(3) model with fitted deformation parameters was used for the calculations of the PSF only in 10 isotopes listed in the table~\ref{tabl1}. The gamma-ray energy intervals near GDR peaks in the tables~\ref{tabl2}-\ref{tabl4} correspond to that used in ref.\cite{Plu2018} for determination of the recommended GDR parameters.

\begin{table*}[htbp]
\center
\caption{{The values of the ratios $R_{\alpha}=<\chi _{\alpha }^{2} >/<\chi _{{\rm SLO}}^{2} >$ of mean values of least-square deviations $<\chi _{\alpha }^{2} >$ in relation to SLO for different PSF models.}}
\label{tabl3}
\begin{tabular}[c]{cccccccc}
\hline\noalign{\smallskip}
Mass numbers of the isotopes  & Gamma-ray energy intervals & SMLO & SMLOe & GLO & TLO(1) & TLO(2) & TLO(3) \\
\noalign{\smallskip}\hline\noalign{\smallskip}
24 $\leq A \leq$ 238 & 5-30 MeV & 0.77 & 0.87 & 1.26 & 4.63 & 4.70 & 2.6 \\
 & Near GDR peaks & 1.08 & 1.67 & 4.12 & 26.15 & 27.66 & 6.08 \\
\noalign{\smallskip}\hline
\end{tabular}
\end{table*}

\begin{table*}[htbp]
\center
\caption{{The values of the ratios $F_{\alpha } =<f_{\alpha } >/<f_{{\rm SLO}} >$ of mean values of rms deviation factors $<f_{\alpha}>$ in relation to SLO for different PSF models.}}
\label{tabl4}
\begin{tabular}[c]{cccccccc}
\hline\noalign{\smallskip}
Mass numbers of the isotopes & Gamma-ray energy intervals & SMLO & SMLOe & GLO & TLO(1) & TLO(2) & TLO(3) \\
\noalign{\smallskip}\hline\noalign{\smallskip}
24 $\leq A \leq$ 238 & 5-30 MeV & 0.96 & 0.98 & 1.04 & 1.13 & 1.10 & 1.10 \\
 & Near GDR peaks & 1.00 & 1.00 & 1.06 & 1.24 & 1.22 & 1.31 \\
\noalign{\smallskip}\hline
\end{tabular}
\end{table*}

It can be seen from tables~\ref{tabl3}-\ref{tabl4}, that calculations using the SMLO model generally better describe of the experimental data then within all other models. For the GLO model, the average ratios $R_{{\rm GLO}} $ in the full interval (5-30 MeV) slightly exceed the unity, so the quality of fitting with this model is comparable to the SLO model. The description of the experimental data using TLO models with standard theoretical values for rigid triaxiality deformation parameters $\beta$, $\gamma$ is worse then for SLO model and also the models with energy-dependent width in approximation of axially deformed nuclei.

\section{Conclusions and discussion}
\label{sec:4}

Based on obtained data, one can conclude that the SMLO model gives better description of the experimental data and can be recommended in the nuclear reaction codes for modeling of the E1 photon strength function for the energy range of 5-30 MeV. For the gamma-ray energies $<$ 30 MeV, the calculations within SMLOe model describe photoabsorption data almost with the same quality like for SMLO. That is no need in separate determination of peak parameters for SMLOe model. For the gamma-ray energies over $\sim$30 MeV, SMLOe model is more preferable because in this model the energy-weighted sum rule is rather good performed.

 Two features must be pointed out. First, the relationship  (\ref{EQ_2}) between cross-section $\sigma _{{\rm E1}} (\varepsilon _{\gamma } )$ (summed over states with all possible total angular momentum) and E1 PSF is, in fact, definition of the E1 PSF determining total photoabsorption cross-section. In nuclear reaction codes for calculation observed characteristics of nuclear reaction, the average photoabsorption cross section $\sigma _{{\rm E1}}^{J} (\varepsilon _{\gamma })$ of a nucleus in the ground state of spin $J_{0} $ with excitation of levels of spin $J$ is used and $\sigma _{{\rm E1}} (\varepsilon _{\gamma } )=\sum _{J=|J_{0} -1|}^{J_{0} +1}\sigma _{{\rm E1}}^{J}  (\varepsilon _{\gamma } )$. In approximation of independency of  $J$ of squares of the reduced matrix elements for dipole transitions \cite{Bar1973}, the partial cross-section  $\sigma _{{\rm E1}}^{J} (\varepsilon _{\gamma } )$ can be presented as product  $\sigma _{{\rm E1}}^{J} (\varepsilon _{\gamma } )=g_{J} \cdot \Phi (\varepsilon _{\gamma } )$ of the statistical factor $g_{J} =(2J+1)/(2J_{0} +1)$ and an independent of $J$ function $\Phi (\varepsilon _{\gamma })$. One can find from formula (\ref{EQ_2}) the following relationship between E1 PSF from (\ref{EQ_3}) and partial photocross-section

\begin{equation}
\label{EQ_26}
\sigma _{{\rm E1}}^{J} =\frac{g_{J} }{3} \, \, \sigma _{{\rm E1}} =\frac{2J+1}{2J_{0} +1} \left(\pi \hbar c\right)^{2} \cdot \varepsilon_{\gamma} \cdot \overrightarrow{f}_{{\rm E1}} (\varepsilon_{\gamma}).
\end{equation}

Second, for some applications (for example, average resonance capture measurements,  astrophysics) it is  needed to calculate photoabsorption cross-section of heated nucleus at some temperature $T$. In this situation, the expression for SLO line-shape (\ref{EQ_5}) is not changed but  expressions for line-shapes GLO and SMLO are changed and their widths $\Gamma _{j}^{\alpha} $ (\ref{EQ_6})-(\ref{EQ_7}) are also modified \cite{RIPL,RIPL2,RIPL3}. New line-shape  ($\bar{F}_{j}^{{\rm SMLO}}$) of the SMLO model equals to product of  the expression (\ref{EQ_5}) and low-energy enhancement factor $L(\varepsilon_{\gamma},T)$,

\begin{equation}
\label{EQ_27}
\begin{array}{l}
{\displaystyle L(\varepsilon _{\gamma},T)\equiv \frac{1}{1-\exp (-\varepsilon _{\gamma } /T)} \, \, \xrightarrow[{\varepsilon _{\gamma } <<\, \, T}]{} \, \, \, \, \frac{T}{\varepsilon _{\gamma } },} \\
{\displaystyle L(\varepsilon_{\gamma},T=0)=1,}
\end{array}
\end{equation}

\noindent i.e.,

\begin{equation}
\label{EQ_28}
\bar{F}_{j}^{\alpha } (\varepsilon _{\gamma } )=L(\varepsilon _{\gamma } ,T)\cdot \frac{2}{\pi } \frac{\, \varepsilon _{\gamma }^{2} \, \Gamma _{j}^{\alpha } }{(\varepsilon _{\gamma }^{2} -(E_{r,j}^{\alpha } )^{2} )^{2} +(\Gamma _{j} ^{\alpha } \varepsilon _{\gamma } )^{2}}.                                                \end{equation}

The new line-shape ($\bar{F}_{j}^{{\rm GLO}}$) of  GLO model consists of two components: a Lorentzian and  additional  term due to Kadmenskij -  Markushev -  Furman (KMF) approach \cite{Sch2007} within Fermi-liquid theory:

\begin{equation}
\label{EQ_29}
\begin{array}{l}
{\displaystyle \bar{F}^{{\rm GLO}} (\varepsilon _{\gamma } )\, =F^{{\rm GLO}} (\varepsilon _{\gamma } )+}\\
{\displaystyle \hspace{0.5 cm} + \frac{\pi }{2} \cdot \varepsilon _{\gamma } \cdot \frac{0.7\, \Gamma _{j}^{{\rm GLO}} (\varepsilon _{\gamma } =0,T)}{(E_{r,j}^{{\rm GLO}} )^{3} }.}
\end{array}
\end{equation}

Note that due to the second term in (\ref{EQ_29}), the energy integrated total photoabsorption cross-section
$\sigma_{{\rm E1}}(\varepsilon_{\gamma})$, (\ref{EQ_29}),  divergates in the temperature-dependent GLO model.

According to the experimental data (\cite{Pan2012,Den1983}, and the references therein) the shape width  in heated nuclei is the temperature-dependent. In the models of SLO, GLO and SMLO, this dependence of the width $\Gamma_{j}^{\alpha}$ is taken into account in form of additional temperature dependent component $\Delta\Gamma_{j}^{\alpha}(T)$:

\begin{equation}
\label{EQ_30}
\Gamma_{j}^{\alpha}=\Gamma_{j}^{\alpha}(\varepsilon_{\gamma},T)=
\Gamma_{j}^{\alpha}(\varepsilon_{\gamma},T=0)+\Delta\Gamma_{j}^{\alpha}(T).
\end{equation}

For GLO and SLO models, the term $\Delta\Gamma _{j}^{\alpha }(T)$ is taken like in the KMF approach \cite{Sch2007}:

\begin{equation}
\label{EQ_31}
\Delta \Gamma _{j}^{\alpha } (T)=g^{\alpha } \cdot 4\pi ^{2} T^{2} ,\, \, \, \, \, g^{\alpha } =\frac{\Gamma _{r,j}^{\alpha } }{(E_{r,j}^{\alpha } )^{2} } .
\end{equation}

The temperature dependence (\ref{EQ_30}) was also adopted for SMLO width in ref.\cite{Gor2019}; the value $g^{{\rm SMLO}} =a/4\pi ^{2} $ with   nuclear level density parameter $a$ was  taken in older variants of the SMLO model \cite{RIPL2,RIPL3}.

For  GDR energies, the expressions (\ref{EQ_29}), (\ref{EQ_30}) correspond to the GDR spreading width within framework of the Fermi-liquid theory in low-temperature limit \cite{Kol1995,Mug2000} with the normalization to the GDR width at zero temperature: $\Gamma _{r} (E_{r} ,T)=g\cdot (E_{r} ^{2} +4\pi ^{2} T^{2} )$, $g=\Gamma _{r} (T=0)/E_{r} ^{2} $. The temperature- dependence of the experimental  GDR widths in warm nuclei can be rather good described by the expressions (\ref{EQ_29}), (\ref{EQ_30}) \cite{Kol1995,Mug2000,Mug2002}. So, the total widths for SLO, GLO and SMLO models are given by the following expressions:  $\Gamma_{j}^{{\rm SLO}} (\varepsilon_{\gamma},T)=g^{{\rm SLO}} \cdot $$((E_{r,j}^{{\rm SLO}} )^{2} +4\pi ^{2} T^{2} ),$ $\Gamma _{j}^{{\rm GLO}} (\varepsilon _{\gamma } ,T)=g^{{\rm GLO}} \cdot $$(\varepsilon _{\gamma }^{2} +4\pi ^{2} T^{2} ),$  $\Gamma _{j}^{{\rm SMLO}} (\varepsilon _{\gamma } ,T)$$=g^{{\rm SMLO}} \cdot $ $(\varepsilon _{\gamma } \cdot E_{r,j}^{{\rm SMLO}} +4\pi ^{2} T^{2} ).$ For extended variants  (SMLOe, GLOe) of the SMLO and GLO models, the expressions for widths are determined by the equation (\ref{EQ_22}). Note that the energy-dependent components of the widths  correspond to transitions of the coherent 1p-1h states generating the GDR to the 2p-2h non-equilibrium states at the excitation energy $U=\varepsilon _{\gamma } $ and $\Delta \Gamma _{j}^{\alpha } (T)$ is connected with transitions to the 2p-2h equilibrium states at different excitation energy  $U\approx a\cdot T^{2}$. The linear dependence on energy of the first component in
$\Gamma_{j}^{{\rm SMLO}} (\varepsilon_{\gamma},T)$ corresponds to the average squared matrix element in the transitions  of the 1p-1h states to 2p-2h with inverse dependence on the energy $\varepsilon _{\gamma } $.

\section*{Acknowledgments}
The authors are very thankful to Stephane Goriely and Tamas Belgya for valuable discussions, and Tamas Belgya for comments on TLO model. This work is partially supported by the International Atomic Energy Agency through a Coordinated Research Project on Updating the Photonuclear Data Library and generating a Reference Database for Photon Strength Functions (F41032).
This is pre-print of an article published in The European Physical Journal A. The final authenticated version is available online at: https://doi.org/10.1140/epja/i2019-12899-6 
%
%

\end{sloppypar}
\end{document}